\documentclass[twocolumn,showpacs,preprintnumbers,amsmath,amssymb]{revtex4}
\usepackage{tabularx,graphicx}\begin{document}
\newcommand{\beq}{\begin{equation}}
\newcommand{\eeq}{\end{equation}}
\newcommand{\beqn}{\begin{eqnarray}}
\newcommand{\eeqn}{\end{eqnarray}}
\newcommand{\bmath}{\begin{subequations}}
\newcommand{\emath}{\end{subequations}}
\newcommand{\rb}{\bar{r}}
\newcommand{\bk}{\bold{k}}
\newcommand{\bkp}{\bold{k'}}
\newcommand{\bq}{\bold{q}}
\newcommand{\bkb}{\bold{\bar{k}}}
\newcommand{\br}{\bold{r}}
\newcommand{\brp}{\bold{r'}}
\newcommand{\vp}{\varphi}

\title{Kinetic energy driven superconductivity and superfluidity }
\author{J. E. Hirsch }
\address{Department of Physics, University of California, San Diego\\
La Jolla, CA 92093-0319}

\begin{abstract} 
The theory of hole superconductivity proposes that superconductivity is driven by lowering of quantum kinetic energy
and is  associated with  expansion of electronic orbits and 
expulsion of negative charge from the interior to the surface of superconductors and beyond. This physics provides a dynamical explanation of the Meissner effect.
Here we propose that similar physics takes place in superfluid helium 4. 
Experimental   manifestations of this physics in $^4He$
are the negative thermal expansion of  $^4He$ below the 
$\lambda$ point and the ``Onnes effect'', the fact that superfluid helium will
creep up the walls of  the  container and escape to the exterior. The  Onnes effect and the Meissner effect are proposed to originate in macroscopic zero point
rotational motion of the superfluids. It is proposed that this physics indicates a fundamental inadequacy of conventional quantum mechanics.

   \end{abstract}
\maketitle 
\section{introduction}

 \begin{figure}
\resizebox{8.5cm}{!}{\includegraphics[width=7cm]{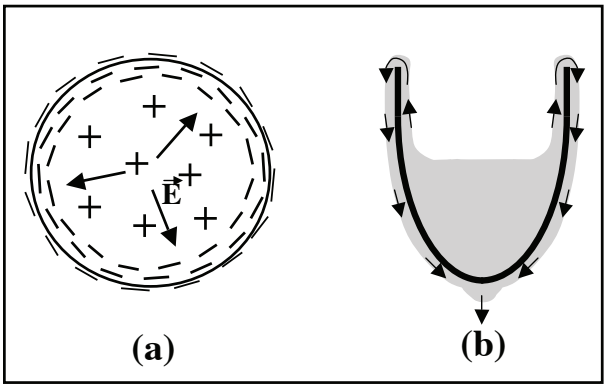}}
\caption { (a)  The superconductor expels negative charge from the interior to the surface. A small amount of charge spills out beyond the surface. 
 (b) Superfluid $^4He$  continuously expels mass from the bulk, that creeps up the inner surface of the container and creeps down the outer surface, at a uniform rate.}
\label{figure1}
\end{figure} 

There exist remarkable similarities in many aspects of the physical behavior exhibited by superfluid $^4He$  and superconductors. This has been  repeatedly found experimentally as well as theoretically and discussed, particularly 
by F. London  and K. Mendelssohn\cite{gorter,london12,mend}, in the early days of research in these fields. 
In this paper we point out that some qualitatively  new physics recently predicted  to
exist in superconductors within  the theory of hole superconductivity also appears to have a remarkable counterpart 
 in superfluid $^4He$.

The theory of hole superconductivity differs in fundamental  ways from conventional BCS-London theory. It proposes\cite{kinetic} that superconductors expel negative charge
from the interior to the surface, a process that is associated with expansion of the electronic wavefunctions and driven by lowering of the quantum
kinetic energy. In the region of excess negative charge that results
within a London penetration depth of the surface a macroscopic spin
current is predicted  to flow in the absence of applied fields, a kind of zero-point motion of the superfluid\cite{sm}. Furthermore the
expelled negative charge is predicted to `seep out' from the surface of the body to the exterior\cite{atom} (see Fig. 1(a)). 

An equivalent physics in superfluid He would correspond to a tendency of the superfluid to expel mass. And indeed such
an effect exists, the `Onnes effect' associated with `Rollin films', shown schematically in Fig. 1(b). Superfluid He will spontaneously creep up the walls of the container,
defying gravity, and escape to the exterior. This effect, first encountered (but not understood)  by Onnes in 1922\cite{onnes1922} and investigated many years later by Rollin\cite{rollin} and especially  by Daunt and Mendelssohn\cite{dm}
in a series of detailed experiments, is arguably  the most remarkable property of superfluid He.
In this paper we propose that the essential physics behind the phenomena displayed in Fig. 1(a) and Fig. 1(b) for superconductors and
superfluids respectively is the same. One difference however is that in the superfluid the mass continuously flows out,
while in the superconductor the expelled negative charge is held back by the strong electric force that results from
the macroscopic charge inhomogeneity.

The current understanding of $^4He$  is principally based on the work of Feynman as described in Ref. \cite{feynman}. 
Unfortunately, at the outset of that review paper Feynman states `We shall omit references to the phenomena involved in the
Rollin film'. Thus we have to focus our attention on earlier theoretical and experimental 
work that properly focused on what we believe is the key physics of 
superfluid $^4He$.

Mendelssohn\cite{mendelssohntransfer,mendelssohntransfer2} and F. London\cite{londontransfer} have pointed out the remarkable similarity in the `transfer phenomena' exhibited by superconductors
and superfluids, namely the flow of mass in superfluid He films depicted in Fig. 1(b)
as well as through capillaries, and the flow of charge within the London penetration depth of the surface that occurs in a superconducting wire carrying a current. They have proposed that the speed at which these
`transfer phenomena' occur is given by a `quantum condition' involving the fundamental constant $\hbar$. For $^4He$, a  similar proposal
had  been made earlier  by Bilj, de Boer and Michels\cite{bilj}. These proposals are not regarded as valid in the contemporary view
of  $^4He$ \cite{atkins,wilks, brewer,tilley}, having been superseded by theories that ascribe the film formation and  flow in $^4He$  exclusively to Van der Waals forces. 
In this paper we argue that the original proposals of Mendelssohn, Bilj et al and F. London were indeed correct, which 
implies that fundamental physics of $^4He$  is missed in the currently accepted understanding of this system.

In addition, being macroscopic quantum systems, superfluids and superconductors provide us with a `window' through which we
can peer into the microscopic world of quantum mechanics and understand it in a new way. In the last sections
of this paper we 
discuss the implications of this physics of superfluids and superconductors
 to the understanding of fundamental quantum mechanics. 

\section{kinetic energy driven superconductivity}

Within conventional BCS theory\cite{tinkham}, carriers lower their potential energy and increase their kinetic energy in the transition to 
superconductivity. Just the opposite is predicted to occur within the theory of hole superconductivity\cite{kinetic}. Experimental evidence for kinetic energy lowering
in the transition to superconductivity in cuprates and pnictides has been seen in optical properties\cite{optical}.

The theory predicts that as the system goes superconducting an $expansion$ of the electronic wavefunction occurs, 
driven by kinetic energy lowering, from
linear extension $k_F^{-1}$ ($k_F=$ Fermi wavevector) in the normal state, to $2\lambda_L$ in the
superconducting state, with $\lambda_L$ the London penetration depth\cite{sm,holecore}. The normal state can be viewed as
electrons residing in non-overlapping `orbits' of radius $k_F^{-1}$, which is of order of the interatomic 
distances since the band is almost full (hole conduction is required). The normal state magnetic
susceptibility is given by the Larmor susceptibility with orbit radius $k_F^{-1}$:
  \beq
 \chi_{Larmor}(r=k_F^{-1})=-\frac{n e^2}{4m_e c^2}k_F^{-2}=-\frac{1}{3}\mu_B^2g(\epsilon_F)
 \eeq
 with $n$ the number of electrons per unit volume,  $g(\epsilon_F)=3n/2\epsilon_F$ the density of states
 at the Fermi energy and $\mu_B=|e|\hbar/2m_ec$ the Bohr magneton, yielding Landau diamagnetism. 
 Because the orbits are non-overlapping, each electron can be at any position in its orbit
 relative to other electrons; there is no `phase coherence'. As the system goes superconducting, the orbits expand
 to mesoscopic radius $2\lambda_L$ and the susceptibility is given by
   \beq
 \chi_{Larmor}(r=2\lambda_L)=-\frac{n e^2}{4m_e c^2}(2\lambda_L)^2=-\frac{1}{4\pi}
 \eeq
(using that $1/\lambda_L^2=4\pi n e^2/m_e c^2$), 
 indicating perfect diamagnetism. Because the orbit radius is now much larger than the interelectronic distances the
 orbits are highly overlapping and long-range phase coherence is required to avoid collisions that would
 raise the energy of the system.
This orbit expansion has associated with it negative charge expulsion from the interior to the surface of
superconductors, as depicted schematically in Fig. 2.

 \begin{figure}
\resizebox{8.5cm}{!}{\includegraphics[width=7cm]{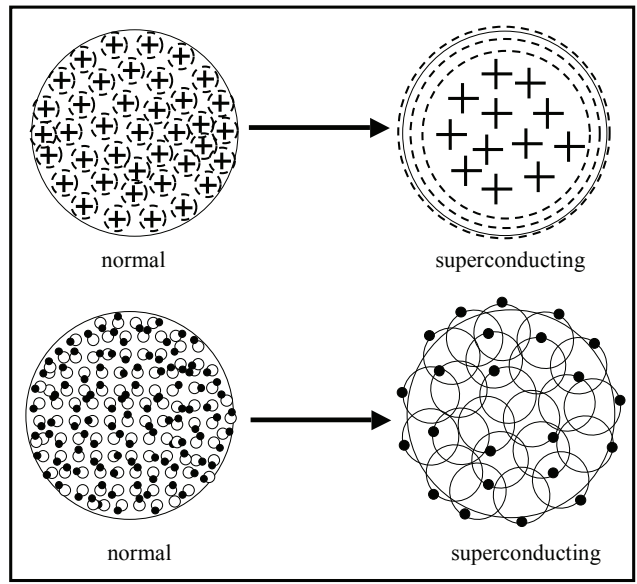}}
\caption {As electrons expand their orbits from radius $k_F^{-1}$ (left lower panel) to radius
$2\lambda_L$ (right lower panel), negative charge moves from the interior to the surface (upper panel). 
The `phase' of each orbit (defined as the position of the electron in its orbit, denoted by the black dot) is random in the normal state because the orbits
are non-overlapping. When the orbits overlap in the superconducting state, the phase of one orbit is correlated to the phase
of all other orbits and there is `macroscopic phase coherence' to avoid collisions.}
\label{figure1}
\end{figure} 

In the $2\lambda_L$ orbits,  opposite spin electrons traverse their orbits in
opposite directions. 
The electrons acquire their orbital speed through the spin-orbit interaction of the magnetic moment of the
electron with the ionic charge background as the orbits expand  from the microscopic scale ($k_F^{-1}$)
($k_F=$ Fermi wavevector)   to $2\lambda_L$, through a `quantum spin Hall effect'\cite{sm}. 
The speed of the electrons is derived from the strength of the spin orbit interaction as given by Dirac's equation 
with the electric field generated by the ionic background and yields\cite{sm,slafes}
\beq
\bold{v}_\sigma^0=-\frac{\hbar}{4m_e\lambda_L}\hat{\sigma}\times \hat{n}      
\eeq
where $\hat{n}$ is the outward normal to the surface.
The orbital angular momentum of electrons in the $2\lambda_L$ orbits is
\beq
L=m_e v_\sigma^0 (2\lambda_L)=\frac{\hbar}{2}   .
\eeq
The superposition of these orbital motions gives rise to a macroscopic spin current flowing 
within a London penetration depth of the surface of superconductors, parallel to the surface\cite{electrospin}.
This is a macroscopic {\it zero point motion} of the superconductor which is predicted to exist even
at zero temperature\cite{sm}.

From Heisenberg's uncertainty principle one infers that a particle confined to a linear distance $\Delta x$ undergoes
zero-point motion with minimum momentum
$p=\hbar/(2 \Delta x)$ and speed $v= \hbar/(2m\Delta x)$, with $m$ the mass. Therefore we can interpret the spin current
speed Eq. (3) as zero-point motion 
originating in the `confinement' of the carriers in a region of width $\Delta x=2\lambda_L$ from the surface.
It is important to note however that the net speed of the superfluid spin current is parallel to the surface
rather than in the `confinement' direction which is perpendicular to the surface.

In the presence of an external magnetic field, electrons in the expanding orbits acquire an additional speed due to the action of the Lorentz force, giving rise 
to a surface current that will screen the magnetic field. Thus, this physics provides a $dynamical$ explanation of the Meissner effect\cite{meissner}.
As discussed in Ref. \cite{sm}, the speed of an electron near the surface in the presence of an applied magnetic field is
\beq
v_\sigma=v_\sigma^0 \pm \frac{e}{m_e c}  B \lambda_L
\eeq
so that it increases (decreases) for spin antiparallel (parallel) to the magnetic field. Superconductivity is destroyed when 
the speed of the electrons with spin parallel to the magnetic field goes to zero, at the critical field given by\cite{sm}
\beq
B_s=\frac{\hbar c}{4|e|\lambda_L^2}.
\eeq
Therefore, electrons carrying the charge current in the presence of a magnetic field have maximum speed $2v_\sigma^0$.
The expression Eq. (6) is essentially $H_{c1}$, the lower critical field of type II superconductors\cite{tinkham},
and the charge current corresponding to half the carriers moving at speed $2v_\sigma^0$ (or all the carriers moving at speed $v_\sigma^0$
in the same direction) gives the critical current according to Silsbee's criterion\cite{tinkham}:
\beq
J_c=n_s e v_\sigma^0=\frac{c}{4\pi\lambda_L} H_{c1}
\eeq
using the well-known expression\cite{tinkham}
\beq
\frac{1}{\lambda_L^2}=\frac{4\pi n_s e^2}{m_e c^2}   
\eeq
with $n_s$ the superfluid density. The excess negative charge near the surface is
$\rho_-=en_sv_\sigma^0/c$, and the critical current Eq. (7) can be understood as arising from 
$\rho_-$ propagating at the speed of light\cite{electrospin}.

\section{kinetic energy driven superfluidity} 

The fact that kinetic energy lowering plays a dominant role in the physics of liquid helium has been known for a long time\cite{simon}.

As discussed by London\cite{londonbook}, the average interatomic distance in liquid helium ($4.0\AA$)  is considerably larger
than what would correspond to the minimum in the Van der Waals potential energy ($3.0\AA$). The molar volume of $^4He$ at zero
pressure is almost three times as large as the volume for which a closed-pack arrangement of He atoms has
minimum potential energy.  The reason that He at low temperatures does not optimize its potential energy is of course
that the zero-point kinetic energy at the distance of minimum potential energy  would be way too high\cite{londonbook}.
If packed at that density, the system will $expand$ from this classically optimal configuration, driven by quantum
kinetic energy lowering, to the actual density that minimizes the sum of potential and kinetic energies. And 
quantum zero-point
motion is of course the reason why helium does not solidify but remains liquid under its own vapor pressure
even at zero temperature, unlike any other substance\cite{simon}.

In addition, when cooled below the $\lambda$ point the system expands further: the thermal expansion
coefficient of helium changes sign from positive above   the $\lambda$ point to negative below\cite{thermalex}.  
Unlike the case of water crystallizing into ice, where
the expansion can be understood from classical geometrical and potential energy considerations, there is no classical explanation for this
behavior for the structureless He atoms. 
The Hamiltonian for $^4He$ is given by
\beq
H=-\sum_i \frac{\hbar^2}{2m_{He}}\nabla^2_i +\frac{1}{2}\sum_{i\neq j}
U(|\vec{r}_i-\vec{r}_j|)\equiv K+U
\eeq
and the energy of the system is the expectation value
\beq
E=<\Psi|K|\Psi>+<\Psi|U|\Psi>
\eeq
with the many-body wavefunction $|\Psi>$. In the normal liquid, the expectation value $<\Psi|U|\Psi>$ is already larger than what it would be for a higher density,
since  the average interatomic distance is  a factor $1.3$ larger than what would correspond to the minimum in the Van der Waals potential energy 
between atoms. When the system is cooled below the $\lambda$ point  the average interatomic distance increases  further, by about
$5$ parts in $1,000$ between $T_\lambda$ and $T=0$. This implies that the average potential energy $<\Psi|U|\Psi>$ increases further, hence the average
kinetic energy $<\Psi|K|\Psi>$ has to decrease. Thus, 
the fact that $^4He$  expands rather than contracts
as the temperature is lowered below the $\lambda$ point 
 is clearly driven by kinetic energy lowering, i.e. `quantum pressure'.
This (experimentally verified)  expansion of the system as it becomes superfluid parallels the (not yet experimentally verified) orbit expansion and charge expulsion that we predict
occurs when a metal enters the superconducting 
state, also driven by kinetic energy lowering\cite{kinetic,emf}. 

However, the kinetic energy lowering is much larger than what would result from this volume expansion.
 In the normal state of $^4He$ the atomic wavefunction is confined to a ``cage'' determined by its neighboring atoms\cite{atkins2},
with zero-point energy  $\sim h^2/8m\delta^2$, with $\delta$ the average interatomic distance. An increase in $\delta$  of
only $5$ parts in $1,000$ cannot of course account for the condensation energy which is of order $k_B T_\lambda$ ($T_\lambda=2.19^oK$), given that 
$h^2/8m\delta^2k_B=3.71^oK$ for interatomic distance $\delta\sim4\AA$. Consequently, in addition to $\delta$ expanding
the atomic wavefunctions must expand
way beyond $\delta$ and strongly overlap with each other, as depicted in the lower right panel of Fig. 2 for superconductors, thus lowering the
quantum kinetic energy by an amount of order $k_BT_\lambda$. 
For superconductors, the transition is associated with expansion of the electron wavefunction from $k_F^{-1}$, the interelectronic distance, to $2\lambda_L$. There is no analog of the $\lambda_L$ length for $^4He$, so it is reasonable to conclude that for $^4He$ the wavelength expands from $\sim \delta$ to the
entire region allowed by geometrical constraints.

Incidentally, negative
thermal expansion has also been proposed to be associated with kinetic energy lowering in metallic ferromagnets\cite{ferromagnets}.

\section{Thickness of $^4He$ films}

The films that $^4He$ forms below the $\lambda$ point on a vertical surface dipped in the fluid are found to be remarkably thick, typically of order $300 \AA$ or $80$ atomic layers
at a height $\sim 1cm$ above the surface. 
In contrast, films at temperatures above the $\lambda$ point are found to be substantially thinner, sometimes as thin  as $10 \AA$\cite{atkins, wilks,brewer,tilley}.

This observation suggest that the thickness of films is intimately tied to the superfluid character. Remarkably however, the generally accepted view is that the
film thickness is determined solely by Van der Waals forces between the $^4He$ atoms and the surface. In a simple description, the equation relating the 
film thickness $d$ to the height above the surface $z$ is taken to be\cite{atkins}
\beq
m_{He} g z -\frac{m_{He}\alpha}{d^3}=0
\eeq
with $g$ the acceleration of gravity and $\alpha$ a constant describing the potential energy of interaction of $He$ atoms and the surface, originating in 
Van der Waals forces. This view implies that the film thickness should be the same below and above the $\lambda$ point, the observation that it is not is
attributed to the different heat conduction properties of $He$ below and above the $\lambda$ point\cite{atkins, wilks,brewer,tilley}.

Instead, the unified physical picture proposed in this paper   suggests that  the driving mechanism for the formation of the thick films below the $\lambda$ point is kinetic energy lowering. This mechanism of film formation was in fact
proposed long ago by Bilj et al\cite{bilj}, but is not considered valid in the contemporary view. The equation relating height $z$ and thickness $d$ is\cite{bilj}
\beq
m_{He} g z -\frac{h^2}{8m_{He}d^2}=0   .
\eeq
The second term originates in the fact that as an atom is added to the film it pays zero-point kinetic energy 
$h^2/8m_{He}d^2$, however the zero-point  energy of all the atoms at that height is lowered by twice that amount due to the slight increase in $d$
in adding the extra atom.

Equations (11) and (12) make qualitatively different predictions on the power-law dependence of film thickness versus height, namely
$d\propto z^{-1/3}$ and $d\propto z^{-1/2}$ respectively. Experiments have not clearly favored one over the other\cite{atkins, wilks,brewer,tilley}. 
The magnitude of the observed thicknessess
is approximately consistent with either explanation. However, the point of view advocated in this paper strongly favors the physics proposed by Bilj et al
as the essential explanation for the formation of thick superfluid $^4He$ films.

\section{transfer phenomena in superfluids and superconductors}

F. London wrote a paper in 1945 entitled ``Planck's constant and low temperature transfer''\cite{londontransfer}. The paper discusses 
in a unified way the 
low temperature transfer   of mass in superfluid He films and  in superconductors carrying a current near the surface.
The same concepts were discussed in the work of Mendelssohn and of Daunt and 
Mendelssohn\cite{mendelssohntransfer,mendelssohntransfer2}.
London defines the mass transfer rate as
\beq
R=\int jdx
\eeq
with
\beq
j=m n v_s
\eeq
  the   mass current density for a superfluid with particles of mass $m$ and number density $n$ moving at average speed
  $v_s$ assumed parallel to the surface, and $x$ the direction perpendicular to the
surface. $R$ gives the mass transferred per unit time along a surface of unit width. 
It has units $g /cm \cdot s$, which is also {\it angular momentum per unit volume}, or angular momentum 
density. Hence $R/n$, with $n$ the superfluid number density, or the number density of superconducting electrons, 
has units of angular momentum. London remarks that the
maximum value of $R/n$ observed experimentally  ($\equiv R_c/n$) $both$ in superfluid $^4He$ and in superconductors,
is of order $\hbar$, more precisely\cite{londonbook}:
\beq
R_c=n\frac{\hbar}{2}
\eeq
where $n$ is the number density of the carriers giving rise to the mass transfer.
For  the superfluid flowing in a layer of thickness $d$, the  transfer rate is
\beq
R=n m v_s d
\eeq
 (where the mass $m$ is $m_e$ for the superconductor, $m_{He}$ for $^4He$), so that Eq. (15) yields
\beq
\frac{R_c}{n}=mv_c d=\frac{\hbar}{2}
\eeq
where we denote by $v_c$ the critical velocity.

For $^4He$ we have
\beq
m_{He}=6.65\times 10^{-24} g
\eeq
and the number density is
\beq
n=2.182\times 10^{22} cm^{-3}   .
\eeq
The thickness of superfluid $^4 He$ films is found to be approximately $300\AA$. The condition Eq. (17) yields for the critical
 transfer speed
\beq
v_c=26.4 cm/s
\eeq
and for the critical mass  transfer rate
\bmath
\beq
R_c=1.15\times 10^{-5} \frac{g}{cm \cdot s} .
\eeq
or for the critical volume transfer rate
\beq
\frac{R_c}{nm}=0.793\times 10^{-4} \frac{cm^3}{cm \cdot sec}
\eeq
\emath
These values are close to what is usually found experimentally\cite{atkins,wilks}. 
Mendelssohn and coworkers have shown in a series of detailed experiments that when different film flow rates are found it can be attributed
to extraneous effects such as impurities adsorbed on the surfaces that change the effective geometrical transfer perimeter\cite{mend1950}.
Daunt and Mendelssohn have furthermore shown\cite{fountain}  that the critical transfer rate is the same when the film flow is driven by heat supplied to 
the container (thermo-mechanical effect) as when it is driven by a gravitational potential difference.

For superconductors we have for  the current density  $J=n_s ev_s$, with $n_s$ the density of
superconducting charge carriers, so that the mass transfer rate is
\beq
R=\frac{m_e}{e}\int Jdx=-\frac{m_e c}{4\pi e}\int  \frac{\partial B}{\partial x} dx=-\frac{m_e c}{4\pi  e}B
\eeq
where $B$ is the magnetic field and we have used $\vec{\nabla}\times\vec{B}=\frac{4\pi}{c}\vec{J}$.
The magnetic field $\vec{B}$ points along the $+\hat{y}$ direction and varies in the $\hat{x}$ direction,
the charge current flows in the $+\hat{z}$ direction and the mass current flows in the $-\hat{z}$ direction,
and the surface is in the $y-z$ plane, as shown in Fig. 3. The critical transfer rate is then
\beq
R_c=-\frac{m_e c}{4\pi e} H_{c1}
\eeq
where $H_{c1}$ is the lower critical field given by Eq. (6):
\beq
H_{c1}=\frac{\hbar c}{4|e|\lambda_L^2}=\frac{\phi_o}{4\pi\lambda_L^2}
\eeq
with $\phi_0=hc/2e$ the flux quantum. Eq. (24) yields a flux $\phi_0$ for magnetic field
$H_{c1}$ through a circle of radius $2\lambda_L$. The critical transfer rate Eq. (23) is then
\beq
R_c=\frac{m_e c^2}{4\pi e^2 \lambda_L^2} \frac{\hbar}{4} .
\eeq
or, using Eq. (8)
\beq
R_c=n_s\frac{\hbar}{4}=n_\sigma\frac{\hbar}{2} .
\eeq
Here, $n_\sigma=n_s/2$ is the density of carriers of one spin only.  Eq. (26) agrees with London's 
formula Eq. (15) {\it if the electrons contributing to the mass current are electrons of one spin orientation only}.

 \begin{figure}
\resizebox{8.5cm}{!}{\includegraphics[width=7cm]{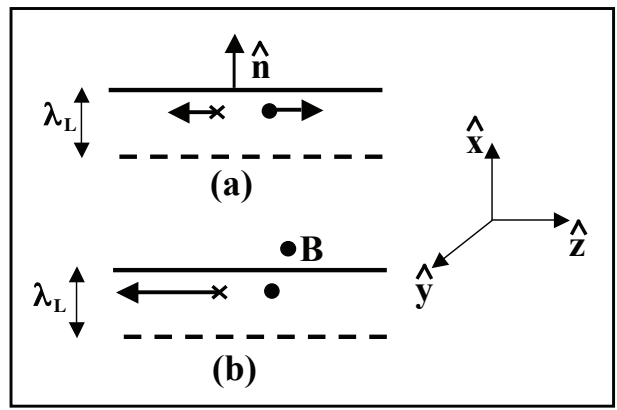}}
\caption {The flow of carriers near the surface of a superconductor is shown. The normal to the surface points in the
$\hat{x}$ direction. In (a) no magnetic field exists and no charge current flows. Electrons with spin pointing
in the $+\hat{y}$ ( $-\hat{y}$) direction (out of and into the paper respectively) move in the $+\hat{z}$ ( $-\hat{z}$) direction with average speed $v_\sigma^0$ (Eq. (3)).  In (b), a magnetic field $B$ of magnitude
equal to the lower critical field $H_{c1}$ exists pointing in the $\hat{y}$ direction. Electrons with
spin pointing into the paper move at average speed $2v_\sigma^0$ and those with opposite spin have stopped. }
\label{figure1}
\end{figure}

This result follows naturally within the theory of hole superconductivity. In the absence of applied magnetic
field, carriers of opposite spin move along the surface in opposite directions with speed given by $v_\sigma^0$ (Eq. (3))
within a layer of thickness $d=\lambda_L$ from the surface, as shown in Fig. 3(a). When the applied magnetic field reaches
the value $H_{c1}$, electrons of one spin orientation come to a stop and those of opposite spin orientation move
at speed $2 v_\sigma^0$, as discussed in Sect. II. The critical transfer rate Eq. (15) is
\beq
\frac{R_c}{n_\sigma}=m_e (2v_\sigma^0) \lambda_L=m_e (2 \frac{\hbar}{4m_e \lambda_L})\lambda_L=\frac{\hbar}{2}
\eeq
in agreement with Eqs. (15) and (26).
Thus, the theory of hole superconductivity allows for a simple interpretation of the London transfer equation Eq. (15).

\section{quantum zero-point   diffusion}

In a superconducting wire fed by normal metal leads, the transport of current occurs with no drop in electric
potential across the superconductor, i.e. in the absence of an accelerating force, as shown schematically in 
Fig. 4(a) . Daunt and Mendelssohn have constructed a very interesting analog of this situation for superfluid
$^4He$, the ``double beaker'' experiment shown schematically in Fig. 4(b)\cite{beaker}. Two concentric vessels
are filled with liquid $He$, and $He$ will flow spontaneously from the inner vessel through the outer vessel
to the outside. The key observation is that the levels of the inner and outer vessels remain identical
throughout the process. Thus, there is no gravitational potential difference between the inner and outer
vessels, hence no gravitational force driving the fluid from the inner to the outer vessel. The transfer of
matter from the inner to the outer vessel without a driving force parallels the transport of current across
the superconducting part of the wire in Fig. 4(a) without electric potential difference between its
ends.

 \begin{figure}
\resizebox{8.5cm}{!}{\includegraphics[width=7cm]{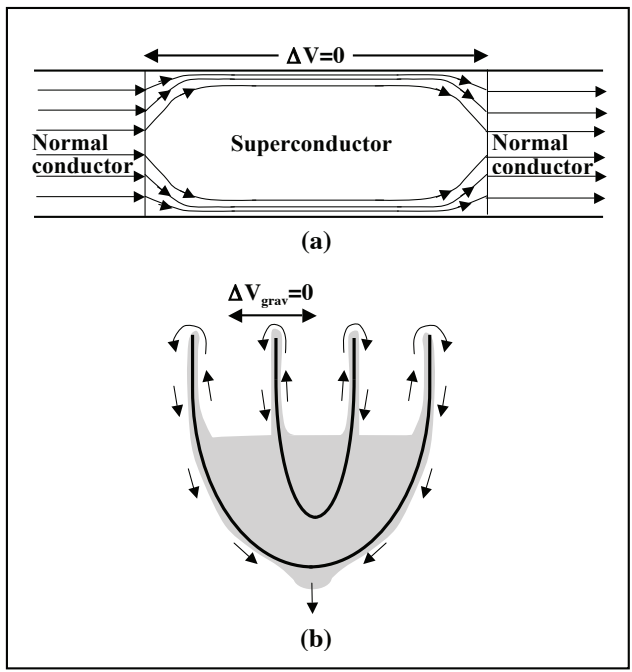}}
\caption {(a) Current flow in a superconducting wire which is fed by normal conducting leads.
 The flow lines are calculated in Ref. \cite{londonbook2}. There is no electric potential difference between both ends
 of the superconductor. (b) Flow of superfluid $^4He$ along surfaces in double beaker experiment. 
There
 is flow of superfluid from the inner to the  outer beaker and from the outer beaker to the exterior, gradually emptying
 both beakers.  However the levels in the inner and outer beaker are always identical throughout this
 process, so there is no gravitational potential difference
between them  ($\Delta V_{grav}=0$).
}
\label{figure1}
\end{figure} 

Mendelssohn has strongly advocated the view that {\it ``the momentum of frictionless transfer is derived from zero-point energy''}
and that these transport  phenomena in superconductors and superfluid $^4He$ 
can only be  explained by {\it ``zero-point diffusion''}\cite{mendelssohntransfer}.
He denotes the superfluid particles as ``z-particles'', and postulates that these transfer processes occur because
the z-particles have zero-point energy, and that the transport processes in the superconductor and liquid helium films 
{\it ``is simply due to the diffusion of z-particles under their zero-point momentum''}.  In his words,
{\it ``If at some place outside (dx), z-particles are removed, zero-point diffusion must take place
and a macroscopic flow of z-particles in the x-direction will occur. This process entails no actual acceleration of
z-particles but merely a greater mean free path in the x-direction.''}

Mendelssohn   emphasizes that the Bose-Einstein gas has $no$ zero-point energy, and therefore it cannot
provide an explanation for this pressure-independent flow. He states  
{\it ``The process of zero-point diffusion which we have introduced in order to explain these phenomena
does not exist in the ideal Bose-Einstein gas.''} Therefore, Mendelssohn concludes
{\it ``According to our considerations, zero-point energy is necessary for the appearance of frictionless transport,
and in fact no condensation to zero velocity takes place in superconductors and probably not in liquid
helium either.''}

In other words, the conventional understanding of superfluidity as described by Bose condensation into a macroscopic
$p=0$ state cannot explain the film flow in superfluid $He$, and the conventional understanding of superconductivity
as a condensate into the conventional BCS state cannot explain the current flow in superconductors.
We fully agree.

 \begin{figure}
\resizebox{8.5cm}{!}{\includegraphics[width=7cm]{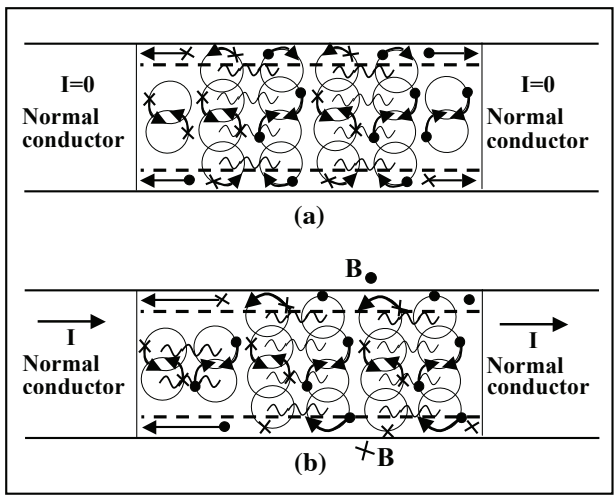}}
\caption {Superconducting wire, some $2\lambda_L$ orbits are shown schematically. The wiggly lines connect Cooper pair partners.
 (a) In the absence of a charge current,
a pure spin current circulates. Near the upper surface, electrons with spin into (out of) the paper circulate to the
left (right), near the lower surface the situation is reversed. There are also vertical spin currents near the
vertical boundaries of the superconductor (not shown). In the interior of the superconductor the spin currents cancel out.
(b) In the presence of a charge current, electrons of spin into
the paper move faster near the upper surface (where the magnetic field points out of the paper), and electrons of spin out of the paper move faster near the lower surface (where the magnetic field points into the paper). There is a net charge
current to the right and a net mass current to the left.
}
\label{figure1}
\end{figure} 

Instead, the ground state of superconductors proposed within the theory of hole superconductivity does
possess  macroscopic zero-point motion, as shown schematically in Fig. 5(a). The ``zero-point diffusion'' in this case
however does result in a change in the speed of the individual electrons when a charge current flows
(Fig. 5(b)). For example, for electrons near the upper surface those with spin pointing into (out of)  the paper 
speed up (slow down). The force responsible for these changes is simply the transient Faraday emf
associated with the changing magnetic field in going from Fig. 5(a) to Fig. 5(b). Nevertheless, confirming
Mendelssohn's intuition, the critical speed associated with the critical transfer rate is found to be closely
related to the speed of the pre-existent zero-point motion Eq. (3).

Thus, following Mendelssohn's reasoning, 
the analogous behavior of $He$ film flow and superconducting charge flow   leads us   to
conclude that {\it superfluid $^4He$ must possess  ground state zero-point motion closely related to the
zero-point motion predicted  in superconductors within the theory of hole superconductivity}.

Conventional quantum mechanics predicts zero-point momentum $p\sim \hbar/d$ {\it perpendicular to the surface} for a film
of thickness $d$ on a  surface. However {\it it does not predict} that this momentum will redirect
itself in direction parallel to the surface.
Instead, in the $2\lambda_L$ orbits predicted  within our model, the magnitude of the
zero-point velocity can be understood with
conventional quantum mechanics as arising from confinement in a surface layer of order of the London
penetration depth and oriented perpendicular to the surface, yet the resulting motion of the spin current is
in direction $parallel$ to the surface\cite{electrospin}. This change of direction is closely associated with
the fact that this zero-point motion is $rotational$. We conclude therefore that {\it $^4He$ must also possess
rotational zero point motion}.

\section{quantum pressure, uncertainty, and rotational zero point motion}

In conventional quantum mechanics there is no ``rotational zero-point motion''. 
For example, a quantum particle in a ring,
whether a fermion or a boson, has a ground state orbital wave function which is constant as function of 
the azimuthal angle and has orbital angular momentum $L=0$ according to the Schrodinger equation.

Instead, we have proposed\cite{double,aromatic}  that electrons in rings have minimum angular momentum $\hbar/2$, and
hence rotational zero-point energy (just like in superconductors according to the theory of hole
superconductivity) originating in an intrinsic double-valuedness of the electron's orbital
wavefunction.

The assumption that electrons have rotational zero-point motion provides a new understanding of
electronic ``quantum pressure''\cite{kinetic,emf}. The kinetic energy of a particle of mass $m$ and orbital angular
momentum $L$ in an orbit of radius $R$ is $L^2/2mR^2$, of the same form as the kinetic energy in
quantum mechanics for $L\sim \hbar$. For $L$ non-zero, reducing the region that the particle occupies
would decrease $R$ and  increase
the kinetic energy and thus provides a $mechanical$ explanation of ``quantum pressure''.
Similarly we can understand Heisenberg's uncertainty principle $\Delta p \Delta x \sim \hbar$ ``mechanically'',
if the left side actually represents an angular momentum of the particle.

 \begin{figure}
\resizebox{7.5cm}{!}{\includegraphics[width=7cm]{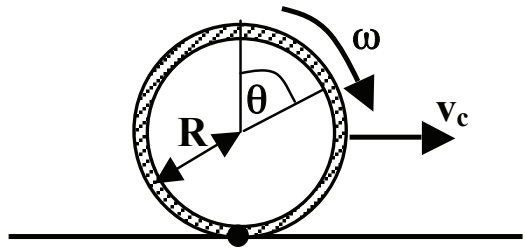}}
\caption {Mechanical picture of a  quantum-mechanical particle propagating with speed $v_c$. It 
has a ``radius'' $R=\hbar/m_e v_c$, 
circumference $\lambda=2\pi R=h/m_e v_c$, rolls without slipping with
angular frequency $\omega=v_c/R$, has linear momentum $p=h/\lambda$,  
kinetic energy $E=\hbar \omega$  and  angular momentum $L=\hbar$.
}
\label{figure1}
\end{figure}

Since  a boson exerts ``quantum pressure'' just like a fermion does and is also subject to Heisenberg's uncertainty principle,
we are led to the conclusion that bosons too must possess rotational zero-point motion, contrary to the predictions of
Schrodinger's equation. We believe that this is
strikingly confirmed by the properties of $^4He$ films discussed in this paper, that vividly display this motion.
More generally, also other  peculiar transport properties of $^4He$ according to Mendelssohn\cite{mend1946} are clear evidence that the superfluid exerts ``zero-point pressure'' that is not accounted for
in the physics of Bose condensation; they would be accounted for by quantum pressure resulting from zero-point rotation, just as we have proposed for
superconductors.

If ``quantum pressure'' and Heisenberg's uncertainty principle are associated with rotational motion, it is natural
to conclude that the quantum ``phase'' of a particle's wavefunction is associated with an azimuthal angle
of rotation. So let us picture a quantum particle of mass $m$ as a ``rim'' of radius R, and assume that
``phase coherent'' propagation corresponds to  {\it rolling without slipping} as depicted in Fig. 6. If the center of mass is moving with
speed $v_c$, the angular velocity for rolling without slipping is
\beq
\omega=v_c/R
\eeq
and the angular momentum is
\beq
L=I\omega=mR^2\omega
\eeq
with $I=mR^2$ the moment of inertia. The linear momentum is given by
\beq
p=mv_c=\frac{L}{R}=\frac{h}{\lambda}
\eeq
if we assume $\lambda=2\pi R$, i.e.  that the circumference of the rim is it's ``wavelength'', and $L=\hbar$ is its
angular momentum. The kinetic energy is the sum of translational and rotational kinetic energies:
\beq
E=\frac{1}{2}mv_c^2+\frac{L^2}{2I}=m\omega^2R^2=L\omega=\hbar\omega=hf
\eeq
with $f=\omega/2\pi$ the frequency of rotation.

Thus, we obtain the well-known de Broglie relations Eqs. (30) and (31), consistent with the laws of mechanics, without assuming that the particle
is a ``wave'' nor the existence of a ``wave packet'' (necessary to account for the factor of $2$ difference between group velocity and phase velocity in the
conventional theory). The radius $R$ is given by $R=\hbar/m_e v_c$ for a particle propagating with speed
$v_c$. If the particle is confined to a region $\Delta x$ this will force $R$ to be smaller than $\Delta x$ and
the frequency of rotation $\omega=\hbar/mR^2$ will increase, thus increasing the kinetic
energy $\hbar \omega$. 

 \begin{figure}
\resizebox{7.5cm}{!}{\includegraphics[width=7cm]{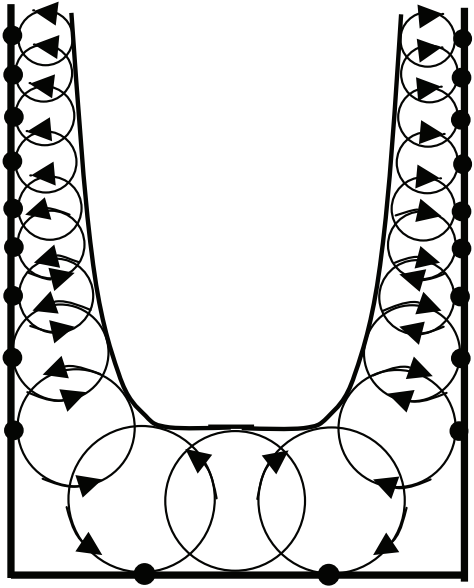}}
\caption {Schematic depiction of $^4He$ in the superfluid phase. The atoms undergo rotational zero-point motion
with radius as large as possible compatible with the geometrical constraints. The film flow against gravity results
from rolling without slipping of the atomic orbits along the vertical surfaces of the container, driven by
the rotational zero-point motion. 
}
\label{figure1}
\end{figure}

In $^4He$, the atoms condense into a coherent wave function describing all the atoms in the condensate.
We assume that  just as in the description of superconductors within the theory of hole superconductivity,
the $^4He$ atoms will undergo macroscopic rotational zero-point motion coherently, as shown schematically in Fig. 7. 
Unlike the case of superconductors where there is the length $\lambda_L$ determined by the mass and
charge of the electron and the density  (Eq. (8)), the length scale of rotational zero-point motion here is presumably determined by the
geometrical constraints and grows as large as possible to minimize the zero-point kinetic energy, as shown
in Fig. 7. The ``rolling without slipping'' motion of the He atom ``rims'' causes the liquid to climb up the walls
of the container, driven by the energy of zero-point motion.
The thickness of the films coating the surface as function of  height can be understood using the
model  of  Bilj et al\cite{bilj}.

Other proposals that revise fundamental aspects of quantum mechanics have been made recently. Nikulov\cite{nikulov} postulates the existence of an azimuthal ``quantum force'' 
to explain the generation of currents in superconducting rings. In contrast to what is discussed here, Nikulov's force would only act when phase coherence is established. 
Hestenes\cite{hestenes} in a series of papers has proposed a rotational ``zitterbewegung'' for electrons that would manifest itself 
``in every application of quantum mechanics''. These proposals may be related to the physics discussed here.

\section{discussion} 

The theory of hole superconductivity predicts that  (1) 
kinetic energy lowering plays a key role in the physics of superconductivity, since superconductivity is driven by
kinetic energy lowering; 
(2)  the wavefunction of the superfluid electrons expands in the transition to superconductivity;
(3) negative charge is expelled from the interior of the superconductor to the surface and beyond; (4)
 macroscopic zero-point motion exists, in the form of a spin current, in the ground state of
superconductors;  (5)  the `Meissner pressure'\cite{londonbook2}  that allows superconductors to expel magnetic fields
in defiance of Faraday's electromotive force is {\it quantum pressure} originating in kinetic energy lowering.
None of these predictions is part of the conventional understanding of superconductivity within
London-BCS theory, and none of these predictions has yet been experimentally verified.

So it is indeed remarkable that each of these predictions has an already $known$ counterpart in the physics of
superfluid $^4He$, namely (following the same numbering as in the previous paragraph):
(1) kinetic energy lowering causes  $^4He$ to expand its volume, from what it would be to optimize the
Van der Waals potential energy, to almost $3$ times larger, and causes $^4He$ to remain liquid under its own
vapor pressure down to zero temperature; (2) in the transition to superfluidity, $^4He$ 
expands further as the temperature is lowered below the $\lambda$ point (negative thermal expansion coefficient  below $T_\lambda$);
(3) mass is expelled from the interior of a $^4He$ container to the exterior (Onnes effect);
(4) macroscopic motion of the superfluid occurs spontaneously in Rollin films; (5) the Rollin film spontaneously
climbs the wall of the container, in defiance of the force of gravity.

This coincidence suggests that (i) superfluidity and superconductivity are even more closely related than 
conventionally believed, and (ii) that  the predictions of the theory of hole superconductivity for superconductors 
are  likely to be  true. In contrast, in the conventional understanding, while many analogies in the behavior of superconductors
and superfluid $^4He$ are clearly recognized\cite{tilley}, kinetic energy lowering plays absolutely no role  in superconductivity. And while several other proposals have been made for
kinetic-energy lowering mechanisms to explain `unconventional' superconductivity in the high $T_c$ 
cuprates\cite{k1,k1pp,k1p,k2,k3,k4,k5,k6,k7,k8,k9,k10,k11,k12,k13,k14,k15}, no connection  between any of  these mechanisms and the physics of superfluid $^4He$ 
has been proposed
so far to our knowledge.

The fact that superfluid $^4He$, a macroscopic quantum system, will spontaneously climb the walls of a container, vividly  suggests
that one is witnessing  a macroscopic manifestation of quantum zero-point motion, i.e. the inability of  quantum
particles to remain at rest. The  fact that the transfer rate per unit density ($R/n$) is of order $\hbar$
lends strong support to this conception. It is really remarkable that   in the generally accepted understanding of $^4He$
quantum zero point motion plays absolutely no role in the creeping up of helium along walls\cite{atkins, wilks,brewer,tilley}.
This is because the conventional understanding of $^4He$ as a Bose condensate does not provide an explanation for it: while
confinement over a film thickness $d$ would raise the zero-point momentum in direction perpendicular to the surface, it would
not give rise to zero-point momentum in direction parallel to the surface.

The properties of superconductors listed in the first paragraph result from the prediction that superconducting electrons undergo
{\it rotational} zero-point motion\cite{sm}. Thus, we are led to the prediction that superfluid $^4He$ atoms also undergo rotational zero-point motion.
The fact that the mass transfer rate per unit number density 
is of order $\hbar$ for both superconducting current and
superfluid $He$ films, observed long ago but not understood, has a simple explanation within the
framework presented here: the critical speed is the speed of macroscopic ground state 
rotational zero-point 
motion in both superconductors and superfluid $^4He$.

This prediction has not yet been verified either for superconductors or superfluid $^4He$. 
However it is well known that rotational motion plays a fundamental role both in superconductors (e.g. vortices in type II materials)
and in superfluid $^4He$ (vortices, ``rotons''). The point of view discussed in this paper suggests that rotational motion in both systems
pre-exists in the ground state as zero-point motion, and is merely made more apparent in excited states.

Beyond superfluids and superconductors, we have proposed that rotational zero-point motion would provide an alternative and more compelling explanation of quantum pressure,
Heisenberg's uncertainty principle, and the stability of matter, than the conventional explanation based on the Schrodinger equation\cite{kinetic}. 
We have shown here that rotating particles {\it rolling without slipping} propagate according to de Broglie's relations.
For electrons, we have proposed  that rotational zero-point motion arises from an intrinsic double-valuedness of the orbital
wavefunction\cite{double,aromatic}, intimately tied to the electron's half-integer spin. However, the considerations in this paper lead to
the conclusion that a system of $^4He$ atoms, bosons, also undergoes rotational zero-point motion. If so, the fundamental origin of
rotational zero-point motion cannot be attributed to half-integer spin. On the other hand it is satisfying that ``quantum pressure'' would have
a unified explanation for both fermions and bosons as arising from rotational   zero-point motion, as is the fact that the
fundamental constant $\hbar$ that determines the behavior of both fermions and bosons in quantum mechanics, has units of angular momentum.

Thus, the considerations in this paper suggest that Schrodinger's equation, as well as Dirac's equation and Klein Gordon's equation from which
Schrodinger's equation derives for fermions and bosons respectively, are  inadequate to provide a general description of
reality, because they do not predict rotational zero-point motion. If so, the correct equations remain to be discovered.


\begin{references}
\bibitem{london12} F. London has written a two-volume book series entitled `Superfluids', where Volume I deals with
superconductors and Volume II with superfluid $^4He$, emphasizing the common aspects of the phenomena
both through the title of the series and throughout its contents.
\end{references}
\end{document}